\documentstyle[aps,psfig,preprint]{revtex}
\begin{document}
\draft
\title{A self-consistent quantal description of high-$K$ states 
\\in  the tilted-axis cranking model}
\author{Makito Oi$^{1,2}$\footnote{m.oi@surrey.ac.uk}, Philip M. Walker$^{1}$, and Ahmad Ansari$^{3}$}
\address{$^{1}$ Department of Physics, University of Surrey, 
Guildford, Surrey GU2 7XH, United Kingdom. \\ 
$^{2}$Department of Applied Physics, Fukui University, 3-9-1 Bunkyo, 
Fukui 910-8507, Japan.\\
$^{3}$Institute of Physics, Bhubaneswar 751 005, India.}
\date{\today}
\maketitle 
\begin{abstract}
A self-consistent and quantal description of high-$K$ bands 
is given in the framework of the tilted-axis cranking model. 
(With a $\theta=90^{\circ}$ tilt angle with respect to $x$-axis,
this cranking model is equivalent to the $z$-axis cranking.)
The numerical results of the HFB calculations in this framework 
are compared with experimental data for two quasi-particle excited bands
with $K^{\pi}=6^+$ in $^{178}$W.

\end{abstract}
\addvspace{7mm}



The nuclear shell structure shows us that
most of nuclei are deformed 
except some doubly and singly closed shell  nuclei and 
a few of their neighbours.
According to microscopic calculations \cite{KB68}, most of the
deformed ground states have prolate shape.
In addition,
the presence of high-$K$ isomers in the well deformed region
($A\simeq 160 \sim 180$) indicate 
axial symmetry is well conserved in many nuclear excited states \cite{WD99}.
Then, $K=\sum_i\Omega_i$ is a good quantum number in a wide range
of nuclear states.
()$\Omega$ denotes the projection of single-particle
spins on the quantization axis, or $z$-axis.)
In such a situation, nuclear collective rotation
is restricted along the $x$-axis, and yrast states
(lowest states for given spin)  are pictured in terms of
one-dimensional static rotations.

The conventional self-consistent (one-dimensional) 
cranking calculations (SCC) 
about the $x$-axis are able to express the  yrast states 
in such a way that the solutions at lower spins 
correspond to the ground-state rotational band (g-band) 
and those at higher spins to the rotation-aligned states (s-band).
Although there are arguments about the validity of 
the cranking model in the band crossing region \cite{ER82,HN90},
SCC works well outside this region and 
is known as a useful tool 
in many microscopic high-spin calculations, in particular, 
in the cases where low-$K$ states are concerned.

In recent studies of high-spin physics,
new types of nuclear rotations have been discovered in experiments,
and explained by theoretical models.
These rotations are not restricted any longer to 
 rotations about the $x$-axis:
for example, rotations of the magnetic dipole moment in
near-magic nuclei (magnetic rotations \cite{Fr93}), 
tilted-axis rotations of triaxial deformed nuclei 
(chiral rotations \cite{DFD00}),
and dynamical two-dimensional rotations in the multi-bands
crossing region (wobbling motions \cite{OAHO00}).
In these phenomena, 
high-$K$ states play significantly important roles, 
especially in the context of tilted rotations.
Microscopic treatments of these phenomena are presented principally
by means of the tilted-axis cranking (TAC) model \cite{Fr93,HO95,MMM00}.
However, as for self-consistent descriptions of 
these multi-quasi particle excited high-$K$ states,
there have been fewer attempts except a well known prescription to
exchange the corresponding Hartree-Fock-Bogoliubov (HFB) solutions
for the second solutions with negative energy \cite{RS80}.

TAC calculations developed by Frauendorf \cite{Fr93}
have been playing important roles to understand several high-$K$
phenomena, but the method is not fully self-consistent because
it fixes the deformation parameters and gap parameters through
the iteration procedure.

Considering the increasing importance of
such high-$K$ states 
it is necessary to search for a tilted-axis version of SCC
for the purpose of physically natural descriptions of high-$K$ states.

At the same time, quantum mechanical treatments 
are required for the description of rotational bands
of multi-quasi particle excited states
because the single-particle (quasi-particle) degree of freedom
has an essential role in these states in addition to
the collective degree of freedom that can be treated in a
semi-classical way.
Angular momentum projection  onto mean-field solutions
is one of the methods which fulfill the requirement.

In a previous paper \cite{OWA01}, we have compared 
numerical results from the self-consistent tilted-axis cranking
calculations with experimental data in $^{182}$Hf, and suggested
that these tilted-rotation minima 
can correspond to the high-$K$ bands as a good approximation.

In this letter,
the above suggestion is examined within the regime of
the tilted-axis cranking model by means of
(1)a mean-field approach based on the
HFB method and (2)a quantal approach based on
angular momentum projection.
Then, attempts are made to describe rotational bands of 
high-$K$ two quasi-particle excited 
states with the tilted-axis cranking model
in a microscopic, fully self-consistent and quantal manner.

First of all, let us present 
our semi-classical picture of high-$K$ bands.
It is accepted that a band head of these bands is produced by 
a two or more quasi-particles excitation that contributes to building
total angular momentum ($I$) along the quantization axis, or $z$-axis.
%
On the other hand, it is assumed that 
rotational members in the high-$K$ bands
build their total angular momentum by collective rotation about the
$x$-axis, as well as single-particle (quasi-particle)
angular momentum along the $z$-axis. 
Therefore, the vector addition of these two angular momenta
gives rise to a total angular momentum vector which is tilted away
from $x$- and $z$-axes. 
%
The tilt angle ($\theta_{\rm cl}$ 
which is defined as an angle of the total angular momentum
from the $x$-axis) can be thus evaluated as 
\begin{equation}
\theta_{\rm cl}^{I=K+\alpha} = \arcsin\left(\frac{K}{I}\right)
=\arcsin\left(\frac{1}{1+\alpha/K}\right),
\label{tcl}
\end{equation}
in which $K$ represents the spin value 
for the band head and $I=K+\alpha$ for 
the $\alpha$-th member in the band.

In our present analysis, we select a high-$K$ band with $K^{\pi}=6^+$ in 
$^{178}$W because it is well studied in experiments \cite{PWD98}.
In $^{178}$W, one $K^{\pi}=6^+$ state is reported and its
configuration assigned as $\nu \left\{ 7/2\left[ 514\right],
5/2\left[ 512\right]\right\}$.


Let us now explain the theoretical frameworks.
The mean-field part is basically the same as the one 
in our previous work \cite{OWA01}:
the Hamiltonian ($\hat{H}$) consists of spherical 
(spherical Nilsson Hamiltonian) and 
residual interaction (the pairing-plus-Q$\cdot$Q force) parts.
The tilted-axis cranked HFB solutions ($\varphi$) 
and corresponding energy ($E(J)$, with $J=(J_1^2+J_2^2+J_3^2)^{1/2}$) 
are obtained by solving the HFB equation by means of the 
gradient method under the constraints 
for particle numbers and angular momentum vectors. 
The tilt angle $\theta$ is introduced in 
the latter constraints as
$\langle\hat{J}_1\rangle = J\cos\theta;
\langle\hat{J}_2\rangle = 0; \langle\hat{J}_3\rangle = J\sin\theta$. 
%
We also constrain the off-diagonal components 
of the quadrupole tensor to be zero, that is,
$\langle\hat{B}_i\rangle = \frac{1}{2}( \langle\hat{Q}_{jk}\rangle 
+ \langle\hat{Q}_{kj}\rangle)=0$,
where $(i,j,k)$ is cyclic. 
Band heads of high-$K$ bands 
are described as solutions for 
$\langle\hat{J}_z\rangle=J=K$ ($\theta=90^{\circ}$).
We call them $z$-axis cranking solutions, 
which are denoted as $\varphi(K)$.
The interaction parameters and configuration space are 
chosen in the same way as in our previous work \cite{OWA01}.
The deformation and gap energy parameters for $J=0$ are
taken from Ref.\cite{KB68}.

The second framework is the quantal part based on the
angular momentum projection,  
which is almost the same as those in Ref.\cite{OOTH98}.
The high-$K$ bands are described as 
the projection from the $z$-axis cranked HFB state ($\varphi(K)$),
\begin{equation}
|\phi^{I=K+\alpha}_M\rangle =
\sum_{K'}g_{K'}^I\hat{P}^{I}_{MK'}|\varphi(K)\rangle,
\label{AMP}
\end{equation}
where $\hat{P}^I_{KK'}=\sum_{q}|IKq\rangle\langle IK'q|$ 
denotes the angular momentum projection operator 
($q$ is a collective symbol for other quantum numbers).
The weight function $g_K^I$ is determined in the course of
diagonalizing the generalized eigenvalue equation to
obtain the projected energy spectrum $\left\{E^I\right\}$
\cite{RS80}. 
On the other hand, the yrast band (corresponding to the g-band at lower
spin)  is described in the same manner
as Ref.\cite{HHR82}. Namely, we perform the angular momentum 
projection for given quantized spin ($I$) 
on the several HFB states with the constraint 
$\langle\hat{J}_x\rangle=J$, and 
obtain the projected states $\phi^I(J)$ with corresponding
energy $E^I(J)$.  
We then choose the members for the yrast band $\phi^I_{\rm y}(J_{\rm min})$ 
satisfying 
$\displaystyle \frac{\partial }{\partial J}E^I(J_{\rm min})=0$.
%

Let us analyze the numerical results for the band heads 
of $K=6^+$ in $^{178}$W, that is, $\varphi(K=6)$.
First, self-consistently  calculated quadrupole deformations 
 
are given: $\beta=0.28$ and $\gamma=0.0^{\circ}$.
That is to say, 
this $z$-axis cranking state possesses well-deformed prolate
shape with almost perfect axial symmetry.
Next, also in the HFB part, we calculate the
contributions to the band head spin 
 from single-particle angular momenta along the $z$-axis.
In our calculations for $^{178}$W, 
the major contributions
are from the neutron negative parity sector ($\nu^-$) 
in the model space. 
About $98\%$ of the total angular momentum of the band head
($J=K=6\hbar$) is produced by spins from
two single-particle orbits; f$_{7/2}$ and h$_{9/2}$, which are consistent
 with the assignment from experiments \cite{PWD98}.

\begin{figure}[h]
\begin{center}
\leavevmode
\parbox{0.9\textwidth}
{\psfig{figure=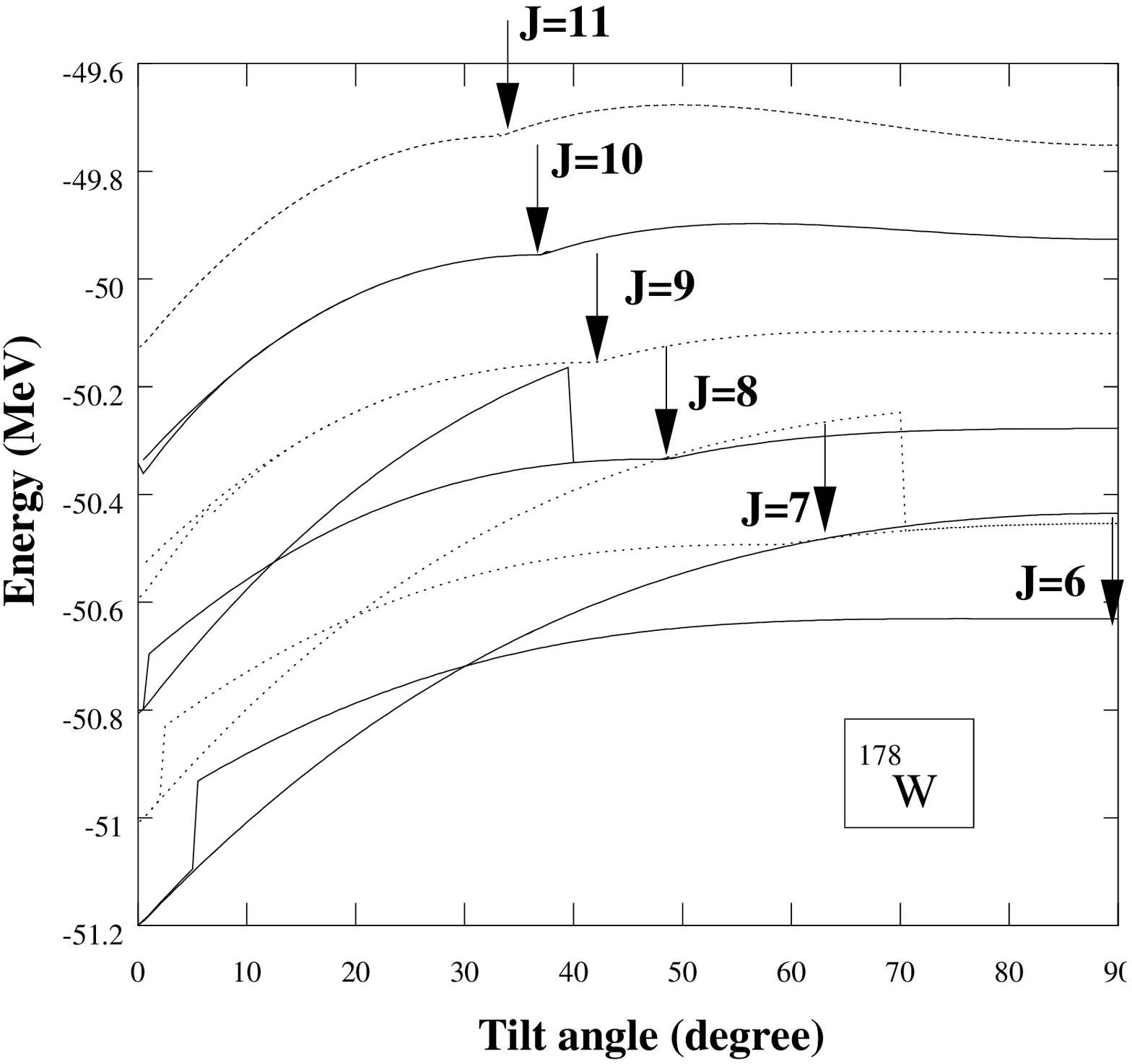,width=0.9\textwidth}}
\end{center}
\caption{Energy curves of $J=6-11\hbar$ 
with respect to $\theta$ for $^{178}$W.
Arrows indicate the position of tilted-rotation minima.}
\label{ene}
\end{figure}

Then, let us go to 
an analysis on the rotational members of high-$K$ bands.
First of all, consider the energy curves with respect to $\theta$, 
which are obtained through the HFB calculations (see Fig.\ref{ene}).
Several local minima are seen in these graphs.L
The lowest ones are seen at $\theta=0^{\circ}$ corresponding to 
the yrast states of g-band character where 
rotations are supposed to be about the $x$-axis.
Other minima are at $\theta=90^{\circ}$, relevant to
high-$K$ states whose spin direction is along the $z$-axis.
If the corresponding spin value is the same as the band head
spin ($K$), these minima can be naturally considered as the band heads
of high-$K$ bands.
The others are found as tilted-rotation minima whose tilt angles
are listed in Table \ref{tlt_cl_tbl}.
Tilt angles for these tilted-rotation minima 
are consistent  with $\theta_{cl}$ in Eq.(\ref{tcl}).
When we look at the $z$-axis components of single-particle spins 
for the tilted-rotation minima,
they are almost the same as those for the band head.
For instance, in the case of a tilted-rotation minimum
at $\theta=38^{\circ}$ for $J=10\hbar$,
the $z$-axis spin components of f$_{7/2}$ and h$_{9/2}$
in the $\nu^-$ sector occupy about $97\%$ of $J_z$ 
$(=J\sin\theta\simeq 6\hbar$).
Furthermore, the distributions of the $x$-axis components of
single-particle spins for these minima
are quite similar to those for the lowest minima at $\theta=0^{\circ}$
(see Table \ref{xalign}).
Thus the tilted-rotation minima
have dual aspects of collective and single-particle rotations
for $x$- and $z$-axes, respectively.
In addition, from calculated deformations for the tilted-rotation minima, 
it can be said that the corresponding
states have well-deformed prolate shape 
($\beta\simeq 0.28$) 
with nearly perfect axial symmetry ($\gamma\alt 0.3^{\circ}$).
Therefore,
this result implies that 
the tilted-axis cranking version of SCC 
can realize rotational members of two quasi-particle excited 
high-$K$ bands as tilted-rotation minima if the proper interaction
is chosen.
%

However, these tilted-rotation minima are expected to be
quite unstable from Fig.\ref{ene}.
Then, it may be necessary to go beyond the mean-field approximation.
As we mentioned above, 
$|\phi^{I=K+\alpha}_M\rangle$ in Eq.(\ref{AMP}) should be employed.

\begin{figure}[h]
\begin{center}
\leavevmode
\parbox{.9\textwidth}
{\psfig{figure=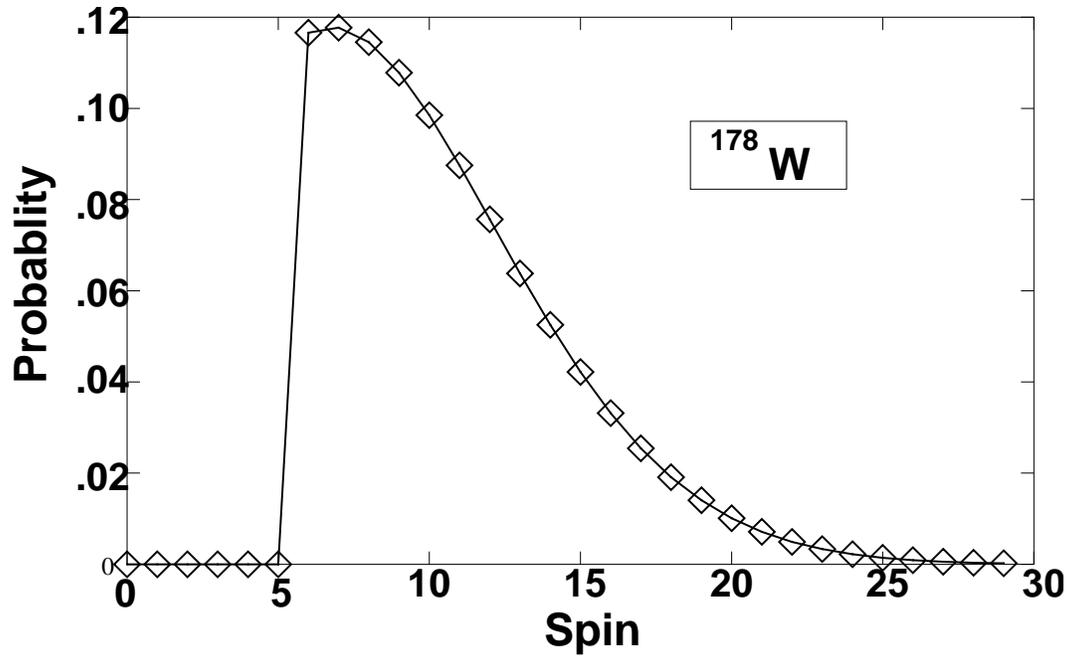,width=0.85\textwidth}}
\end{center}
\caption{Probability distribution $W^I$ for the $z$-axis 
cranked HFB state of $^{178}$W which correspond
to the band head $I=K=6\hbar$. The sum rule for the probability
is satisfied to $99.9\%$ up to $I=29\hbar$.}
\label{AMP_I}
\end{figure}

It is worth looking at angular momentum projection
analysis of the band head states $\varphi(K)$.
In general, a wave packet breaking 
the rotational symmetry is expressed as 
$\displaystyle \sum_{IK}C^I_K|IK\rangle$.
Thus, in our case,  a probability $|C^I_{K'}|^2$
tells us how much the rotational
symmetry is broken by the $z$-axis cranking term.
Fig.\ref{AMP_I} presents the probability distribution for given $I$,
that is, $W^I=\sum_{K'}|C^I_{K'}|^2$.
Due to  signature symmetry breaking by the $z$-axis cranking
term, odd-spin components are equally contained in the HFB state
as well as even-spin components. This feature is satisfactory
for a description of $\Delta I = 1$ bands including high-$K$ bands.
A more important feature is 
a suppression of lower-spin components ($I< 6\hbar$).
This feature reflects that the high-$K$ components from the 
quasi-particle excitation ($K=6\hbar$) are exclusively obtained 
through the $z$-axis cranking term in a self-consistent manner.
%
Although higher-spin components remain as a result of symmetry breaking,
this feature cannot be a problem for 
a description of high-$K$ rotational bands 
because the nuclear structure is preserved adiabatically within the 
band. In fact, this feature is rather preferable for our 
quantal expression (Eq.\ref{AMP}) because 
sufficient of high-spin components are included in these
HFB states.
The probability distribution for given $I$ ($W^I_K=|C^I_K|^2$)
shows the major component in $K$-quantum number.
We have checked for $6\hbar \le I \le 28\hbar$ that
the major component is $K=6\hbar$ and that the ratios 
$\displaystyle W^I_{K=6}/\sum_{K=-I}^{I}W^I _K$
are always more than $99.99\%$.
From the results above, we can say that not only that 
our quantal approach works quite well
to build almost pure $K=6\hbar$ intrinsic states for given $I$,
but also that the mean-field approach is already very good 
in terms of producing states having good (high) $K$-quantum number.

Finally, we compare the energy spectrum for the high-$K$ bands
of $K^{\pi}=6^+$ in  $^{178}$W obtained by 
the tilted-axis cranking version of SCC and the angular momentum
projection. 
Normalizations for the HFB energies are made with respect to
the energy for $J=0$ ( i.e., $E(J)-E(0) \rightarrow E(J)$ ), 
while for the projected energies normalization is made
with respect to the energy at $I=0$ projected from the HFB
state with $J=0$ ( i.e., $E^I(K) - E^{0}(0) \rightarrow E^I$ ).

\begin{figure}[h]
\begin{center}
\leavevmode
\parbox{0.95\textwidth}
{\psfig{figure=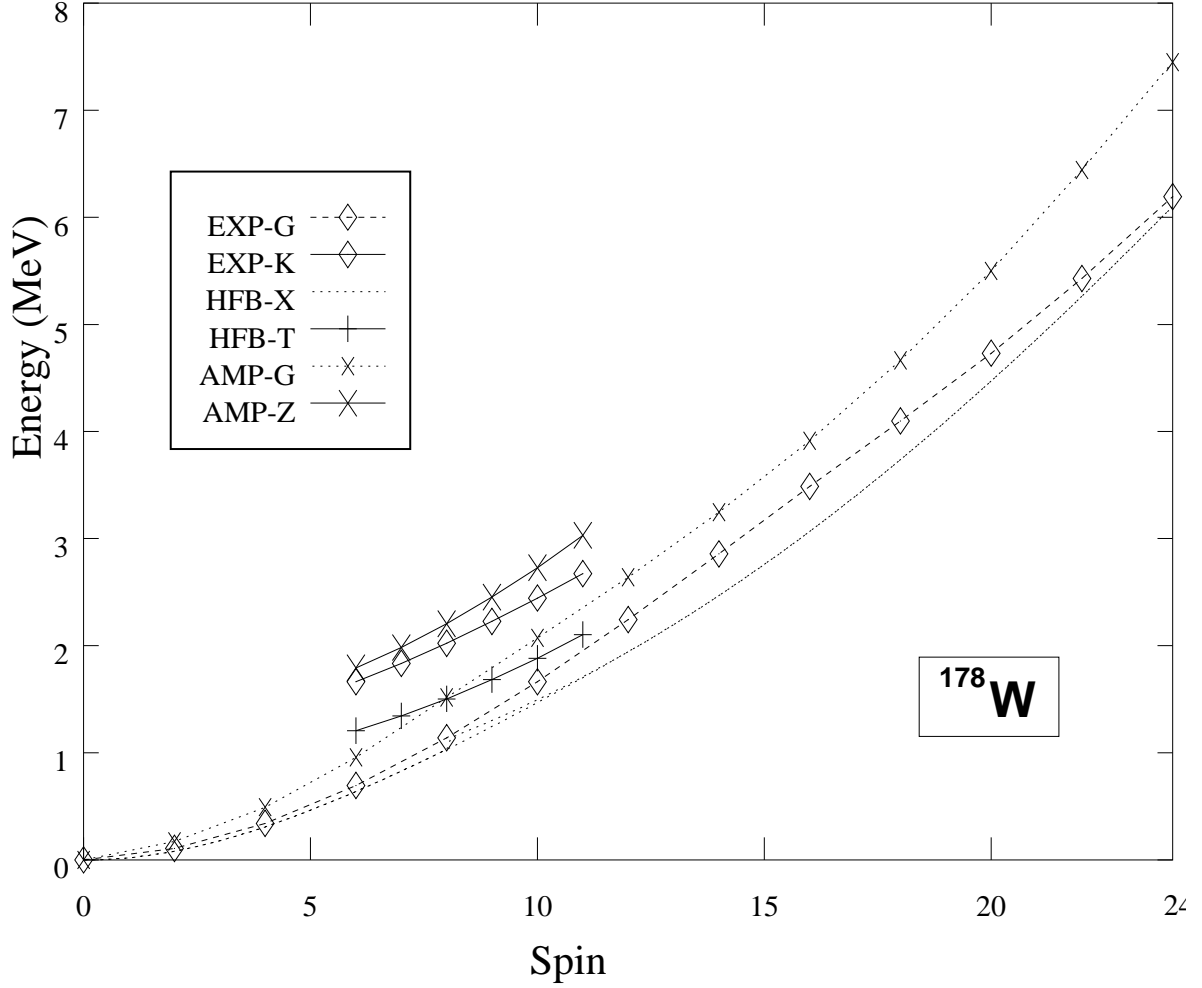,width=0.95\textwidth}}
\end{center}
\caption{Energy spectrum for $^{178}W$. ``Exp-G'' and ``Exp-K''
denote experimental data for the yrast and $K^{\pi}=6^+$ bands,
respectively. For the latter, energies are partially shown up to
$I=11\hbar$. ``HFB-X'' and ``HFB-T'' denote 
the one-dimensional cranked HFB energies 
and tilted-rotation minima in the
tilted-axis cranked HFB calculations, respectively.
``AMP-Z'' (``AMP-X'') corresponds to the projected energy from the
$z$- ($x-$) axis cranked HFB state with $K=J=6\hbar$ ($K=0,J=6\hbar$)}
\label{EI}
\end{figure}

The energies for the band head ($I=K=6\hbar$)
are given as 1.67 MeV (experiment), 1.21 MeV ($z$-axis cranked HFB),
and 1.79 MeV (angular momentum projection), respectively.
As for the rotational members, it is convenient to look at
Fig.\ref{EI}. In spite of the phenomenological 
Pairing-plus-Q$\cdot$Q interaction, the projected energies show
very good agreement with experimental data.
The energies for the tilted-rotation minima
reproduce the experimental data to some extent 
(errors are roughly 400 keV).

The discrepancy of the projected energy 
of the yrast band at high spin 
could be due to the problem in the cranking model, which cannot describe
the band crossing phenomenon properly. 
However, this deficiency does not influence the description 
of the high-$K$ band because it does not meet any band crossing,
at least, about $I=11\hbar$.
Another reason can come from the fact that we have adjusted the residual
interaction for  the mean-field calculations, but not 
for the projection procedure.
This influence is shown in the systematic errors, 
that is, projected energies for both the yrast and high-$K$ bands 
tend to overestimate the experimental values.

In summary, within the tilted-axis cranking model,
we have investigated self-consistent and quantal descriptions
of high-$K$ two-quasi-particle excited bands.
The mean-filed approach shows results consistent with
a semi-classical picture of high-$K$ rotational bands,
and through the projection analysis it is found  that 
the mean-field approach has an effect of angular momentum  projection  
suppressing the irrelevant lower-spin components in the 
symmetry breaking HFB solutions.
On the other hand, 
the quantal approach presents a satisfactory description
in terms of restoration of broken rotational symmetry, 
and can produce more elaborate states taking into account 
higher-order correlations which are 
beyond the mean-field approach.
According to our numerical calculations,
we can conclude that both of the descriptions are good
not only qualitatively but also  quantitatively, at least,
within some limitations owing to the interaction we chose 
in this study.

As to the future work,
we should examin the present method in other situations, 
for instance, four quasi-particle excitations, proton quasi-particle
excitations, negative parity high-$K$ states, and so on.

After a confirmation that the method can describe more general high-$K$
states like the above,
it will be interesting to apply  methods to predict
features for unknown multi-quasi-particle excited high-$K$ bands
such as energies and configurations for the band head,
rotational properties of the bands, etc.
We expect that these methods are quite helpful 
in experimental searches for new high-$K$ isomers in, 
say, the $A\simeq 190$ neutron-rich region that has
 just being explored recently
\cite{ZRP00}.

We thank Dr. S. Frauendorf for useful discussions.
Parts of the numerical calculations 
are carried out at the DEC (Compaq) alpha 
workstation in the center for nuclear sciences (CNS),
the university of Tokyo, and
at the vector parallel super-computer 
Fujitsu VPP700E/128 in RIKEN.
M.O. acknowledges support from
 Japan society of the promotion of sciences (JSPS).

\narrowtext
\narrowtext

\begin{table}[hbt]
\caption{Tilt angles for given spin $I=K+\alpha$:
$\theta_{\rm cl}$ (Eq.(\ref{tcl})) and 
$\theta_{\rm HFB}$ for $^{178}$W (Fig.\ref{ene}).
The band head spin value is $K=6\hbar$.}

\begin{center}

\begin{tabular}{ccc}
{$\boldmath \alpha$} & 
{$\boldmath \theta_{\rm HFB}$}   &  
{$\boldmath \theta_{\rm cl}$} \\
\hline
0 & 90.0$^{\circ}$ & 90.0$^{\circ}$ \\
1 & 58.5$^{\circ}$ & 59.0$^{\circ}$ \\
2 & 48.0$^{\circ}$ & 48.6$^{\circ}$ \\
3 & 42.0$^{\circ}$ & 41.8$^{\circ}$ \\
4 & 38.0$^{\circ}$ & 36.9$^{\circ}$ \\
5 & 33.5$^{\circ}$ & 33.1$^{\circ}$ 
\end{tabular}
\label{tlt_cl_tbl}
\end{center}
\end{table}

\begin{table}[hbt]
\caption{Single-particle spins along the $x$-axis for $^{178}$W 
at $J=10\hbar$. Percentages are for the ratio of the total
spin component along the $x$-axis ($J_x=J\cos\theta$).
The symbols in the first line indicate the isospin and parity, 
For example,  $\pi^+ (\nu^-)$, means the proton positive-parity 
sector (neutron negative-parity sector).}
\begin{center}
\begin{tabular}{cccccr}
$\theta$&
$\pi^+$ & $\pi^-$ & $\nu^+$ & $\nu^-$ & $J_x=J\cos\theta$\\
\hline
$0^{\circ}$ & 0.6 ($6\%$)& 0.7 ($7\%$)  & 4.7 ($47\%$)& 4.0 ($40\%$)&10.0 ($100\%$)\\
$38^{\circ}$& 0.4 ($5\%$)& 0.6 ($7.5\%$)& 4.0 ($50\%$)& 3.0 ($37.5\%$)&8.0 ($100\%$)
\end{tabular}
\label{xalign}
\end{center}
\end{table}


\end{document}